\documentclass[aps,
               pra,
               twocolumn,
               showpacs,
               amsmath,
               amssymb,
               superscriptaddress,
               reprint,
               10pt,
              ]{revtex4-2}

\usepackage{graphicx}
\usepackage{dcolumn}
\usepackage{bm,amssymb,amsmath,dsfont}   
\usepackage{lipsum}
\usepackage{physics}
\usepackage[T1]{fontenc}
\usepackage{txfonts}
\usepackage{bbold}
\usepackage{float}
\usepackage{hyperref}
\hypersetup{
    colorlinks=true,
    linkcolor=blue,
    citecolor=blue,
    urlcolor=blue}

\usepackage{pstricks}
\usepackage{tikz}
\usepackage{color}
\usepackage{xcolor}
\graphicspath{{./}}
\DeclareFontFamily{OT1}{cmm}{}
\DeclareMathAlphabet{\mathcm}{OML}{cmm}{m}{it}
\usepackage{xr}
\newcommand{\eq}[1]{(\ref{eq:#1})}
\newcommand{\Eq}[1]{Eq.\,\eqref{eq:#1}}

\newcommand{\Fig}[1]{Fig.~\ref{fig:#1}}
\newcommand{\fig}[1]{\ref{fig:#1}}

\newcommand{\App}[1]{App.~\ref{app:#1}}

\renewcommand{\i}{\text{i}}

\definecolor{applegreen}{rgb}{0.55, 0.71, 0.0}
\definecolor{byzantine}{rgb}{0.74, 0.2, 0.64}

\makeatletter
\let\cat@comma@active\@empty
\makeatother

\begin{document}

\preprint{APS/123-QED}

\title{Universal dynamics of rogue waves in a quenched spinor Bose condensate}

\author{Ido Siovitz}
\author{Stefan Lannig}
\author{Yannick Deller}

\author{Helmut Strobel}
\author{Markus K. Oberthaler}
\author{Thomas Gasenzer}
\email{t.gasenzer@uni-heidelberg.de}
\affiliation{Kirchhoff-Institut f\"{u}r Physik, 
	Ruprecht-Karls Universit\"{a}t Heidelberg, 
	Im Neuenheimer Feld 227, 
	69120 Heidelberg, 
	Germany}

\date{\today}


\begin{abstract}
Isolated many-body systems far from equilibrium may exhibit scaling dynamics with universal exponents indicating the proximity of the time-evolution to a non-thermal fixed point.
We find universal dynamics connected with the occurrence of extreme wave excitations in the mutually coupled magnetic components of a spinor gas which propagate in an effectively random potential.
The frequency of these rogue waves is affected by the time-varying spatial correlation length of the potential, giving rise to an additional exponent $\delta_\mathrm{c} \simeq 1/3$ for temporal scaling, which is different from the exponent $\beta_V \simeq 1/4$ characterizing the scaling of the correlation length $\ell_V \sim t^{\,\beta_V}$ in time.
As a result of the caustics, i.e., focusing events, real-time instanton defects appear in the Larmor phase of the spin-1 system as vortices in space and time. 
The temporal correlations governing the instanton occurrence frequency scale as $t^{\, \delta_\mathrm{I}}$.
This suggests that the universality class of a non-thermal fixed point could be characterized by different, mutually related exponents defining the evolution in time and space, respectively.
Our results have a strong relevance for understanding pattern coarsening from first principles and potential implications for dynamics ranging from the early universe to geophysical dynamics and micro physics.
\end{abstract}

\maketitle


\emph{Introduction}. 
The study of quantum dynamics far from equilibrium has been of particular interest in recent years.
The time evolution of a system on its way to equilibrium is a rich play-field upon which one can examine various dynamical characteristics, including, e.g., 
prethermalization \cite{Berges:2004ce,Langen:2016vdb}, 
wave turbulence \cite{Zakharov1992a,Nazarenko2011a},
superfluid turbulence \cite{Vinen2006a,Tsubota2008a}, 
and self-similar spatio-temporal scaling dynamics at a non-thermal fixed point \cite{Berges:2008wm,Schole:2012kt,Orioli:2015dxa,Chantesana:2018qsb.PhysRevA.99.043620,Mikheev:2018adp}.
During recent years, studies of such phenomena have intensified, in experiment
\cite{%
Henn2009a.PhysRevLett.103.045301,
Gring2011a,
AduSmith2013a,
Langen2015b.Science348.207,
Navon2015a.Science.347.167N,
Navon2016a.Nature.539.72,
Rauer2017a.arXiv170508231R.Science360.307,
Gauthier2019a.Science.364.1264,
Johnstone2019a.Science.364.1267,
Eigen2018a.arXiv180509802E,
Prufer:2018hto,
Erne:2018gmz,
Navon2018a.doi:10.1126/science.aau6103,
Glidden:2020qmu,
GarciaOrozco2021a.PhysRevA.106.023314}
and theory
\cite{%
Kodama:2004dk,
Barnett2011a,
Marcuzzi2013a.PhysRevLett.111.197203,
Langen2013a.EPJST.217,
Nessi2014a.PhysRevLett.113.210402,
Gagel2014a.PhysRevLett.113.220401,
Bertini2015a.PhysRevLett.115.180601,
Babadi2015a.PhysRevX.5.041005,
Buchhold2015a.PhysRevA.94.013601,
Berges:2008sr,
Nowak:2012gd,
Hofmann2014a,
Maraga2015a.PhysRevE.92.042151,
Williamson2016a.PhysRevLett.116.025301,
Williamson2016a.PhysRevA.94.023608,
Bourges2016a.arXiv161108922B.PhysRevA.95.023616,
Chiocchetta:2016waa.PhysRevB.94.174301,
Karl2017b.NJP19.093014,
Schachner:2016frd,
Walz:2017ffj.PhysRevD.97.116011,
Schmied:2018upn.PhysRevLett.122.170404,
Schmied:2018mte,
Mazeliauskas:2018yef,
Schmied:2018osf.PhysRevA.99.033611,
Williamson2019a.ScPP7.29,
Schmied:2019abm,
Gao2020a.PhysRevLett.124.040403,
Wheeler2021a.EPL135.30004,
Gresista:2021qqa,
RodriguezNieva2021a.arXiv210600023R,
Preis2023a.PhysRevLett.130.031602,
Liu:2022rss,
Heinen:2022rew,
Heinen2023a.PhysRevA.107.043303}, 
many of them in the field of cold gases.

The concept of non-thermal fixed points aims at generalizing upon the description and classification of critical physics in and near equilibrium \cite{Hohenberg1977a,Janssen1979a,Diehl1986a,Janssen1992a} to quenched systems far from equilibrium.
The approach of the system to a non-thermal fixed point is reflected by the self-similar spatio-temporal scaling of the order-parameter correlations \cite{Orioli:2015dxa,Chantesana:2018qsb.PhysRevA.99.043620,Mikheev:2018adp}.
For example, in coarsening and phase ordering kinetics \cite{Bray1994a.AdvPhys.43.357,Puri2019a.KineticsOfPT,Cugliandolo2014arXiv1412.0855C}, the emergence of scaling evolution is, generically, associated with non-linear and topological excitations emerging in a system during its ordering evolution.
The dynamics of such excitations gives rise to a characteristic length scale in the system, which then typically changes in time according to a power law, $\ell_{\mathrm{\Lambda}}(t)\sim t^{\,\beta_\Lambda}$, with a universal scaling exponent $\beta_\Lambda$.

Caustics, the phenomenon of dynamical wave focusing in random media \cite{Berry1977a,Berry1977b,Mumford2017a.JPB50.044005,Kirkby2019a.PhysRevResearch.1.033135,Kirkby2022a.PhysRevResearch.4.013105}, may lead to the formation of wave events of extreme amplitude known as (linear) rogue or freak waves \cite{Metzger2014a.PhysRevLett.112.203903}. 
When propagating in a random medium, the flow of waves branches \cite{Kaplan2002a.PhysRevLett.89.184103,Metzger2010a.PhysRevLett.105.020601,Metzger2013a.PhysRevLett.111.013901,Metzger2014a.PhysRevLett.112.203903,Degueldre2016a.NatPhys12.259}, resulting in the repetitive occurrence of new rogue waves.
Theoretical investigations of this phenomenon connect spatial and temporal scales, making it an appropriate framework for investigating spatio-temporal scaling phenomena also in non-linear media \cite{Arecchi2011a.PhysRevLett.106.153901,Ying2012a.JGeophysR117,Onorato2013a.PhysRep528.47,Green2019a.NJP21.083020,Degueldre2016a.NatPhys12.259} in which rogue waves are known to occur as specific non-linear solutions \cite{Akhmediev2010a.EPJST185.1,Bludov2009a.PhysRevA.80.033610}. 
This framework is of great significance in, e.g., the formation of tsunamis in the ocean and structure formation in the early universe.

\emph{Main result}. 
We consider a one-dimensional spin-1 Bose gas, in which the interactions give rise to a variety of non-linear and topological excitations. 
We study the spatio-temporal pattern of excitations of the Bose fields $\Psi_{m_\mathrm{F}} = |\Psi_{m_\mathrm{F}}|\exp(\i\varphi_{m_\mathrm{F}})$ of the ${m_\mathrm{F}}=0, \pm 1$ magnetic components in the $F=1$ hyperfine manifold, after a sudden quench from the polar into the easy-plane phase \cite{Kawaguchi2012a.PhyRep.520.253}. 
Rogue-wave textures are observed to occur as the result of caustics \cite{Kaplan2002a.PhysRevLett.89.184103,Metzger2010a.PhysRevLett.105.020601,Metzger2013a.PhysRevLett.111.013901, Metzger2014a.PhysRevLett.112.203903, Degueldre2016a.NatPhys12.259} of phase excitations, which effectively propagate in a disordered medium with time-varying correlations formed by the respective other components.
We find the mean time $t_\mathrm{c}$ between caustics to grow in time, $t_\mathrm{c} \sim t^{\, \delta_\mathrm{c}}$, with $\delta_\mathrm{c}= 0.332(3)$, which is distinctly different from the coarsening of the infra-red (IR) length scale of the system, which scales with $\beta_\Lambda\simeq 1/4$ as a result of the non-linear interactions, cf.~\cite{Schmied:2018osf.PhysRevA.99.033611}. 
This leads to the scaling of the correlation length  $\ell_V(t)\sim t^{\, \beta_V}$ of the effectively random potential $V(x,t)$ giving rise to the caustics.
Analysing the generation of caustics in a dynamically coarsening random potential we predict the exponents to be related by $\delta_\mathrm{c} = 4\beta_V/3$.

At the focal point of a caustic, a strong phase kink occurs, which can drive the Larmor phase $\varphi_\mathrm{L} = \varphi_1 - \varphi_{-1}$ into the next Riemann sheet, causing a winding-number jump and with this a real-time instanton event.
These topological defects are robust, allowing us to differentiate them from the background, making them a reliable probe for the dynamics of the system.
We identify these topological defects in the Larmor phase, which take on the form of space-time vortices. 
The characteristic frequency with which such events occur is found to decay with a power law in time, $\Gamma \sim t^{\,-\delta_\mathrm{I}}$, with $\delta_\mathrm{I} = 0.34(1)$, corroborating the temporal behavior of caustics.
The spatial probability distribution function (PDF) of the corresponding kinks decays exponentially with the distance between them, with the mean distance increasing as 
$\expval{r}(t)\sim t^{\,\beta_\mathrm{I}}$,
with the exponent
$\beta_\mathrm{I}=0.26(1)$,
which is consistent with the length-scale coarsening exponent $\beta_\Lambda$ of the order parameter.
\begin{figure*}[t]
    \includegraphics[width=0.85\textwidth]{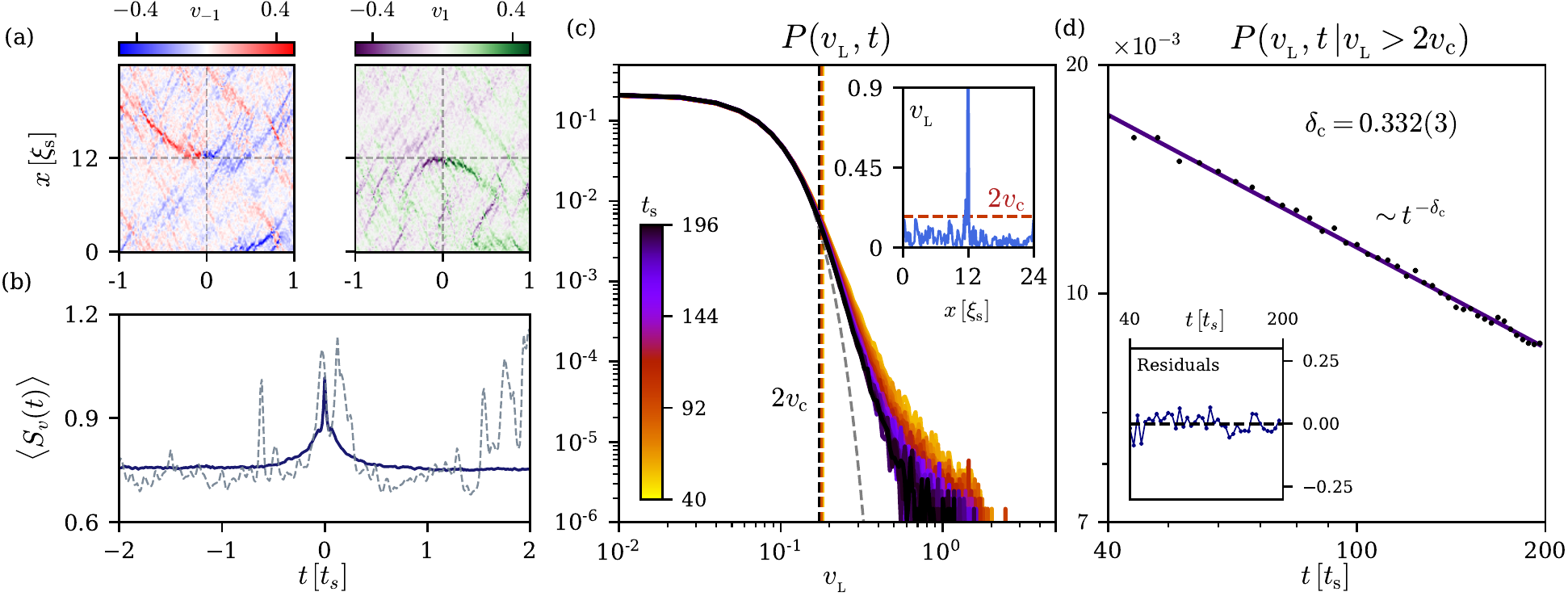}
    \caption{Characteristics of caustics in the system after a quench. Units are set by the spin healing length $\xi_\mathrm{s} = (2Mn|c_1|)^{-1/2}$ and its corresponding spin-changing collision time $t_\mathrm{s} = 2\pi/(n|c_1|)$.
    (a) Excerpt of the space-time evolution of phase defects in the system.
    The phase gradients $\mathcm{v}_{1}=\partial_{x}\varphi_1$ (purple to green) and $\mathcm{v}_{-1}$ (red to blue) show the formation of rogue-wave-like excitations in the condensate which focus on a singular point marked by the cross.
    (b) The scintillation index $S_{\mathcm{v}}(t)$, \Eq{scintillationindex}, 
    around a rogue wave at $t=0$.
    The blue solid line shows the scintillation profile averaged over $\sim 10^3$ (not normalized) rogue waves.
    The dashed line depicts $S_{\mathcm{v}}(t)$ for the single truncated Wigner run in (a).
    (c) PDF of the local Larmor velocity $ \mathcm{v}_{_\mathrm{L}} = \partial_x \varphi_\mathrm{L} = \partial_x (\varphi_1 - \varphi_{-1})$ for different times. The PDF takes the form of a Rayleigh exponential distribution (grey dashed line fit) with a heavy tail. The extreme events are characterized as those with an amplitude larger than $2\mathcm{v}_\mathrm{c}$, where $\mathcm{v}_\mathrm{c}$ is the scale velocity, representing the mean of the upper tertile of events.
    The inset shows that $\mathcm{v}_{_L}>2\mathcm{v}_c$ at the focusing time $t=0$.
    (d) The probability of finding an extreme event as a function of time. A power law decay $t^{-\delta_\mathrm{c}}$, with $\delta_\mathrm{c} = 0.332(3)$ is found. The inset shows the deviation of the fit from the data divided by the data point error.}
    \label{fig:caustic}
\end{figure*}


\emph{The one-dimensional spin-1 Bose gas}
is described by the Hamiltonian
\begin{align}
    H=\int \dd{x} \left[
    \vb*{\Psi}^\dagger \left(-\frac{1}{2M}\pdv[2]{}{x}
    +qf_z^2\right)\vb*{\Psi} +\frac{c_0}{2}n^2 
    + \frac{c_1}{2}\abs{\vb*{F}}^2\right]
    \,,
    \label{eq:Spin1Hamiltonian}
\end{align}
where 
$\vb*{\Psi} = (\Psi_1,\Psi_0,\Psi_{-1})^T$
is the three-component bosonic spinor field and $M$ is the atomic mass. 
Density-density interactions are described by the term $c_0n^2$, where 
$n=\vb*{\Psi}^\dagger\!\cdot\! \vb*{\Psi}$
is the total density. Spin changing collisions are contained in the term $c_1\abs{\vb*{F}}^2$, with
$\vb*{F}=\vb*{\Psi}^\dagger \vb*{f} \vb*{\Psi}$
and $\vb*{f}=(f_x,f_y,f_z)$ are the generators of the $\mathfrak{so}(3)$ Lie algebra in the three-dimensional fundamental representation, cf.~\App{TWA}   \Eq{su3generators}.
$q$ determines the quadratic Zeeman field strength, which causes an effective shift in the energies of the $m_\mathrm{F}=\pm 1$ components relative to the $m_\mathrm{F}=0$ component.
The linear Zeeman effect is transformed away by considering a rotating frame of reference, i.e., absorbed into the time evolution of the fields. 


\emph{Simulations of the dynamics after a quench}.
We consider quenches from the polar 
($c_{1}<0$, $q>2n\abs{c_1}$)
to the easy-plane phase
($c_{1}<0$, $0<q<2n\abs{c_1}$),
where we expect the spin degrees of freedom to be dominantly oriented in the $F_{x}$-$F_{y}$-plane,  see ~\App{GroupTheory} for details.
We prepare the condensate in the mean-field polar phase, which is characterized by a full macroscopic occupation of the $m_\mathrm{F}=0$ component 
$\psi_0(x)/\sqrt{n}=\expval{\Psi_0}/\sqrt{n}=1$,
while the $m_\mathrm{F}=\pm1$ magnetic levels are empty.
The simulations are performed in an experimentally realistic parameter regime for $~^{87}\mathrm{Rb}$, i.e., $|c_1|\ll c_0$.
We add noise to the Bogoliubov modes of the initial state and quench the quadratic Zeeman shift through the second-order phase transition to a final value of $q_\mathrm{f} = 0.9\,n|c_1|$. 
We propagate this state by means of Truncated-Wigner simulations with periodic boundary conditions, see ~\App{TWA} for more details.
Following the quench, instabilities lead to a fast build-up of strong excitations in the relative phases between the different magnetic components, which reflect spatial redistributions of bosons under the interaction-induced constraint of a nearly constant total density $n$, i.e., due to $|c_1| \ll c_0$  \cite{Schmied:2018osf.PhysRevA.99.033611}.
\begin{figure*}[t]
   \centering
    \includegraphics[width=0.9\textwidth]{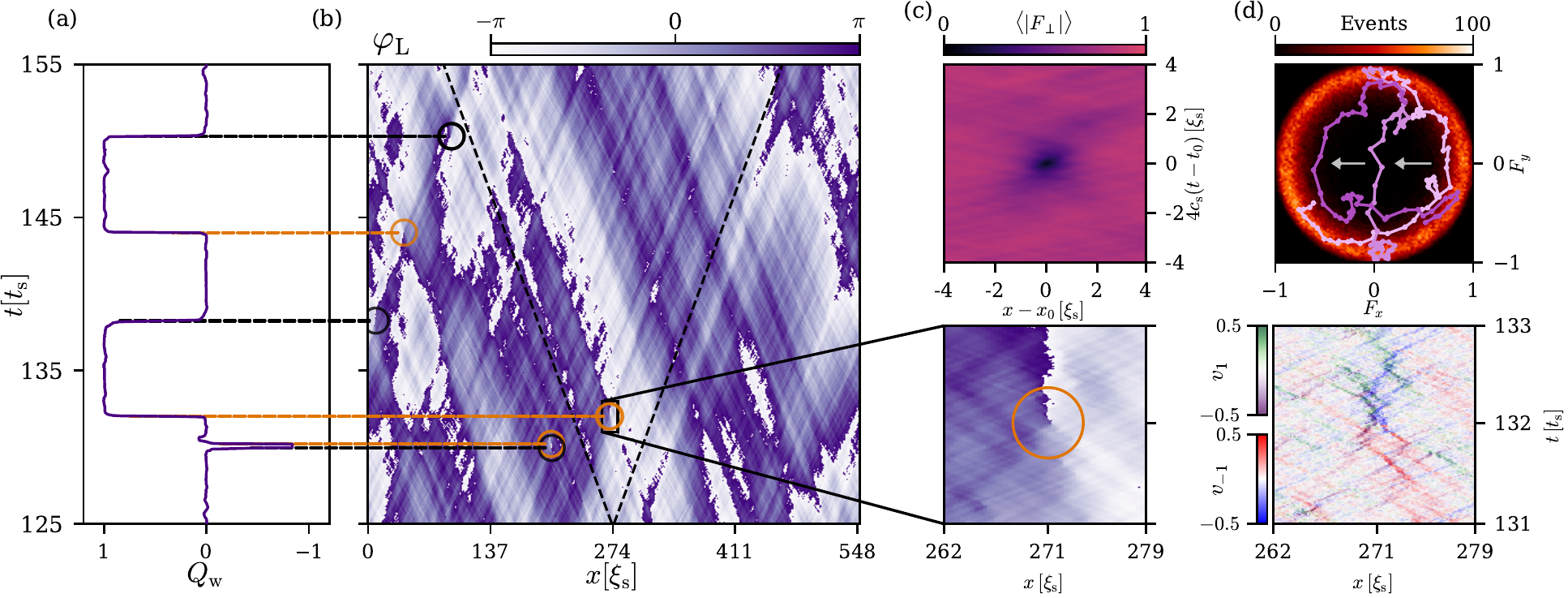}
    \caption{Structures and defects in the time evolution of the Larmor phase after a quench (units chosen as in \Fig{caustic}).  
    (a) Time evolution of the winding number $Q_\mathrm{w}$ for the run shown in panel (b).
    (b) Space-time evolution of the Larmor phase of the transversal spin 
    $F_\perp=\abs{F_\perp}\exp[i\varphi_\mathrm{L}]$
    across the entire system in a single Truncated-Wigner (TW) run, with the spin speed of sound $c_\mathrm{s} = \sqrt{n|c_1|/2M}$ (dashed line).
    In the strongly fluctuating system, vortex structures in space and time are observed, as the phase wraps around one point (cf.~zoom in panel (c)). 
    Instantons (orange) and anti-instantons (black), each cause an integer jump in the winding number $Q_\mathrm{w}(t)$.
    (c) Structure of the real-time instanton. In the upper panel, the averaged $|F_\perp|$ profile of a defect located at $x_0$ at time $t_0$ is depicted.
    The lower panel shows the vortex-like nature of the defect in more detail, around which the Larmor phase winds by $2\pi$.
    (d) The lower panel shows the corresponding intersection of two rogue waves in $\mathcm{v}_{\pm1}$ at the position of the instanton, recall \Fig{caustic}a.
    The upper panel exhibits the temporal evolution (bright to dark pink) of the $F_\perp$ field configuration in spin space, within the window shown in the lower panels.
    The outer circle represents a histogram (black to bright red color code) of spin orientations in the $F_x$-$F_y$ plane averaged over 100 TW runs. 
}
    \label{fig:realspacedefect}
\end{figure*}

\emph{Caustics}. 
The highly excited system in its post-quench time evolution is observed to generate focusing of the magnetic excitations into momentaneous rogue waves in the $m_\mathrm{F}=0$ density, giving rise to density dips in the $m_\mathrm{F}=\pm1$ modes, and thus to rogue-wave-like peaks in the velocity fields $\mathcm{v}_{m_\mathrm{F}} \sim \partial_x \varphi_{m_\mathrm{F}}$ in \Fig{caustic}a (cf. \App{NumericalResults}, ~\Fig{MagneticCaustics}). 
These rogue waves can be characterized as caustics \cite{Kaplan2002a.PhysRevLett.89.184103,Metzger2010a.PhysRevLett.105.020601,Metzger2013a.PhysRevLett.111.013901, Metzger2014a.PhysRevLett.112.203903, Degueldre2016a.NatPhys12.259}, which are signaled, by the scintillation index 
\begin{align}
    S_{\mathcm{v}}(t) = \frac{\langle|\mathcm{v}_{_\mathrm{L}}|^2\rangle_x}{\expval{|\mathcm{v}_{_\mathrm{L}}|}_x^{2}} - 1
    \label{eq:scintillationindex}
\end{align}
as rare extreme events in the velocity fields, where 
$\expval{\cdots}_x$
denotes the spatial average and 
$ \mathcm{v}_{_\mathrm{L}} = \partial_x \varphi_\mathrm{L} = \mathcm{v}_1 - \mathcm{v}_{-1}$.
At times where the system shows strong phase kinks, we expect a strong sudden rise in the scintillation index (see \Fig{caustic}b). 

To study the coarsening dynamics of caustics, we investigate the probability distribution function (PDF) of velocities associated with caustics, which is known to be long-tailed \cite{Safari2017a.PhysRevLett.119.203901,Mathis2015a.SciRep5.12822,Dudley2014a.NaturePhot8.755,Solli2007a.Nature540.1054}, as is confirmed by our simulations, see \Fig{caustic}c. 
The PDF of velocities follows a heavy-tailed Rayleigh exponential form (cf.~\App{Caustics}, \Fig{scintindexandlogv}), implying that the dynamics are driven by coherent wave packets \cite{Metzger2013a.PhysRevLett.111.013901}.
One obtains a scale \emph{velocity of significant waves}, $\mathcm{v}_\mathrm{c}$, as the mean of the upper tertile of the PDF. 
The criterion for rogue waves is then chosen to include those with an amplitude $\mathcm{v}_{_\mathrm{L}}>2\mathcm{v}_\mathrm{c}$ \cite{Dudley2014a.NaturePhot8.755}. 
\Fig{caustic}d shows that the probability of such rare events to occur decays with a power law, $P(\mathcm{v}_{_\mathrm{L}},t|\mathcm{v}_{_\mathrm{L}}>2\mathcm{v}_\mathrm{c})\sim t^{-\delta_\mathrm{c}}$, with $\delta_\mathrm{c}=0.332(3)$.

The underlying time scale of caustic focusing and appearance of rogue or freak wave excitations can be described in the framework of a stochastic non-linear Schr\"odinger equation (NLSE) \cite{Ying2012a.JGeophysR117,Onorato2013a.PhysRep528.47,Green2019a.NJP21.083020}.
To investigate the temporal behavior of extreme events in our system, we consider the equations of motion for $\vb*{\Psi}$,
\begin{align}
  \i \partial_t \vb*{\Psi} = \left[-\frac{\partial_x^2}{2M} + qf_z^2 + c_0n + c_1\vb*{F}\cdot \vb*{f} \right]\vb*{\Psi}
  \,.
  \label{eq:GPE1}
\end{align}
Due to the strong density-density interactions and the disordered behavior of the spin-changing term, the last term of \Eq{GPE1} can be considered as a fluctuating weak random potential 
$V(x,t) \equiv c_1 \vb*{F}(x,t)\cdot\vb*{f}$
added to a NLSE. 
Our numerical simulations show that $\langle V\rangle=0$, since, in the mean over many realizations, the SO$(2)$ symmetry is restored in the easy-plane, and $\langle F_z \rangle=0$. Yet, we obtain exponential correlations in the diagonal elements $\langle \mathrm{Tr}[V(x,t)V(0,0)]\rangle=V_{0}^{2}\exp\big[-x/\ell_V(t)\big]$, with strength $V_{0}$ and a correlation length scale $\ell_V$, whereas the off-diagonal elements of the correlation vanish, see  ~\App{Caustics} for details.

For a propagation in random media, the time needed for the waves to focus, i.e., the \textit{mean time to caustics} $t_\mathrm{c}$, depends only on the correlation length $\ell_V$ of the random medium and on the strength $V_{0}$ of the fluctuations \cite{Kaplan2002a.PhysRevLett.89.184103,Metzger2010a.PhysRevLett.105.020601,Metzger2013a.PhysRevLett.111.013901, Metzger2014a.PhysRevLett.112.203903, Degueldre2016a.NatPhys12.259}. 
In contrast to the standard case studied in the context of caustics, the intricate non-linear interactions between the components of the condensate cause the correlation length to dynamically scale in time. 
Our numerical simulations confirm the scaling of the correlation length of the noise term in \Eq{GPE1} as $\ell_V(t) \sim t^{\,\beta_{V}}$, with $\beta_{V}=0.252(3)$, cf. \App{FtransverseEvolution} and \App{Caustics}.
Generalising the arguments used in \cite{Kaplan2002a.PhysRevLett.89.184103} to Bogoliubov modes, one finds the mean time to caustics to scale as
$t_\mathrm{c}\sim \ell_V^{\,4/3}$,
for details see \App{Caustics}.
Thus, a temporally growing correlation length 
$\ell_V(t)\sim t^{\,\beta_{V}}$,
with
$\beta_{V}\simeq1/4$,
implies that the mean time to caustics scales in time as 
$t_\mathrm{c} \sim t^{\, \delta_\mathrm{c}}$,
with $\delta_\mathrm{c} =4\beta_{V}/3 \simeq 1/3$.

The observed power-law coarsening indicates a close connection with the spatio-temporal scaling of the structure factor $S_{F_{\perp}}(t,p)=\langle F_{\perp}(t,p)^{\dagger}F_{\perp}(t,p)\rangle$ of the transverse spin $F_\perp \equiv F_{x}+\i F_{y} =\abs{F_\perp}\exp[\i\varphi_L]$ as found in \cite{Schmied:2018osf.PhysRevA.99.033611} (cf. ~\App{FtransverseEvolution}),
\begin{align}
    S_{F_{\perp}}(t,p) = (t/t_\mathrm{ref})^\alpha f_\mathrm{s}([t/t_\mathrm{ref}]^{\,\beta} p)
    \label{eq:StructureFactorScaling}
    \,.
\end{align}
Here $f_\mathrm{s}$ is a universal scaling function, which depends only on the momentum $p$, $t_\mathrm{ref}$ is a reference time within the scaling interval, and the scaling exponents $\alpha=0.27(6)$ and $\beta=0.25(4)$ are, within errors, related by $\alpha=d\beta$, $d=1$, ensuring the momentum integral over $S_{F_{\perp}}(t,p)$ to be conserved. 
The scaling is a manifestation of the coarsening of an infra-red (IR) correlation scale, growing as $\ell_{\Lambda}\sim t^{\,\beta_\Lambda}$, which in turn gives rise to the algebraic increase of the noise correlation length scale $\ell_V$ with the same exponent, $\beta_{V}\simeq\beta_\Lambda$, as is confirmed within errors by our simulations.

\begin{figure*}[t]
    \centering
    \includegraphics[width=0.85\textwidth]{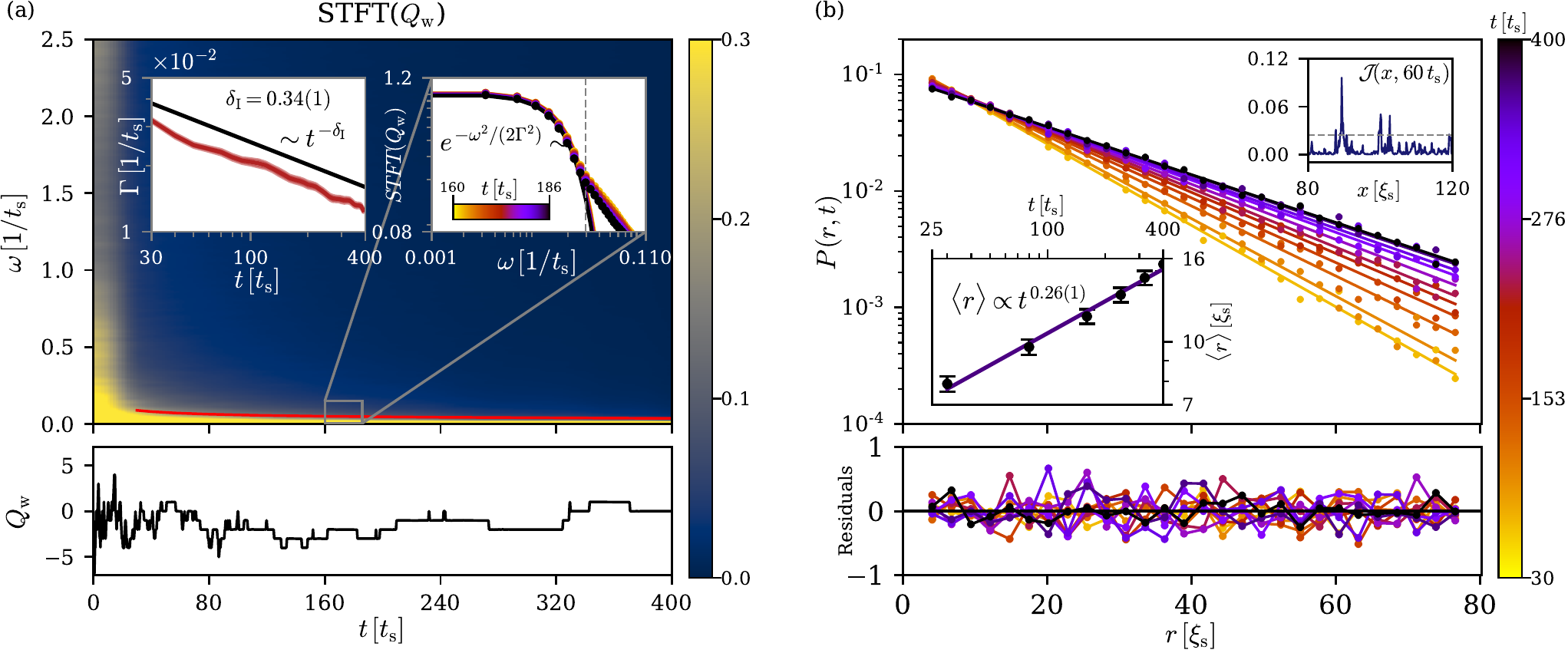}
    \caption{Statistics of the instantons after a quench. 
    (a) Short-time Fourier transform (STFT) of the winding number $Q_\mathrm{w}(t)$ (main panel, color scale), exhibiting a Gaussian fall-off for small frequencies (up to the grey dashed line in the right inset)
    STFT$[Q_\mathrm{w}](t)\sim{\exp}\big\{-\omega^{2}/[2\Gamma^{2}(t)]\big\}$
    (right inset), with width decreasing as 
    $\Gamma(t)\sim t^{-\delta_\mathrm{I} }$, $\delta_\mathrm{I} =0.34(1)$
    (left inset and red line in main panel), confirming power-law coarsening dynamics of the governing timescale of $Q_\mathrm{w}$.
    The lower panel shows the time evolution of the winding number $Q_\mathrm{w}$ for a single realization.
    (b) The PDF of spatial defect separation is found to fall off exponentially,  $P(r,t)\sim A(t)\exp[-r/\zeta(t)]$, with mean distance $\expval{r} (t) \sim t^{\,\beta_\mathrm{I}}$, increasing from early times (yellow) to later times (black), exhibiting power-law coarsening with exponent $\beta_\mathrm{I}=0.26(1)$ (lower inset).
    The upper inset shows the chosen threshold of the current $\mathcal{J}(x,t) = |\partial_x \varphi_\mathrm{L}|\cdot(\expval{|F_\perp|}_x-|F_\perp|)$ for defect detection which corresponds to the top decile of amplitude. 
    The lower panel shows the difference of the data to the fit function divided by the standard deviation of each data point.
    }
    \label{fig:defectseparation}
\end{figure*}

\emph{Real-time instantons in the Larmor phase.} 
In the emerging post-quench dynamics, the confluence of rogue-wave excitations in the velocity fields $\mathcm{v}_{\pm 1}$ manifests itself as an interplay of strong phase kinks in $\Psi_{\pm 1}$.
Analyzing the propagation of the velocity fields over many realizations reveals that the encounter of two focused waves with opposite signs, each in a different component (\Fig{caustic}a), results in an overall phase jump in the Larmor phase, forcing $\varphi_\mathrm{L}$ into the next Riemann sheet. 
As a result, the Larmor phase changes its overall winding number across the system (\Fig{realspacedefect}a, b), an event which we refer to as a \textit{real-time instanton}.
Instantons are of strong relevance in fundamental studies of quantum field theory and matter \cite{Seiberg:1994rs,Seiberg:1994aj}, 
as well as various applications, including false vacuum decay \cite{Hawking1982a.PhysRevD.26.2681,Braden2015a.JCAP03.007,Bond2015a.JCAP09.004}.
Phenomena closely related to the real-time instantons we study here include coherence vortices \cite{Simula2011a.PhysRevA.84.052104} and phase slips \cite{Buechler1999a.cond.mat.11301,Danshita2012a.PhysRevA.85.023638,Danshita2013a.PhysRevLett.111.025303,Errico2017a.PTRSA375.20160425,Polo2019a.PhysRevLett.123.195301}.

%
As can be seen in the lower panel of \Fig{realspacedefect}c, a vortex-type defect occurs in the Larmor phase, at a time $t\simeq132\,t_\mathrm{s}$ and position $x\simeq271\,\xi_\mathrm{s}$, at the intersection of phase kinks, where a strong rogue-wave excitation occurs. 
In \Fig{realspacedefect}d (upper), the instanton defect is seen to result from the field configuration crossing in time the center of the transversal spin plane, causing a local spin length reduction.
As a result, the phase wraps into the next Riemann sheet, giving rise to a change of the overall winding number of the Larmor phase,
\begin{align}
    Q_\mathrm{w} 
    = \frac{1}{2\pi}\int_0^\mathcal{L} \dd{x} \,\partial_x \varphi_\mathrm{L} \in \mathbb{Z}
    \,,
\end{align}
where $\mathcal{L}$ is the system's length.
With the help of a plaquette algorithm correlating jumps in the Larmor phase and dips in the spin length, we localize the instantons in space and time.

During the evolution of the system following the quench, the density of (anti-)instantons decreases, and the probability of the system producing a topological defect reduces as it attempts to settle to a state with constant winding number, see the lower panel of Fig.~\fig{defectseparation}a.
The robustness of these topological defects enables us to distinguish them from the background.
To extract the instanton probability decay, we perform a short-time Fourier transform (STFT) of $Q_\mathrm{w}(t)$ over time windows of width $\Delta t_\mathrm{STFT}=70\,t_\mathrm{s}$.
The resulting STFT$[Q_\mathrm{w}](t,\omega)$ is shown in \Fig{defectseparation}a.
At each time, the winding number jump frequencies display an approximate Gaussian fall-off ${\exp}\{-\omega^{2}/[2\Gamma^{2}(t)]\}$, with scale $\Gamma$ which is extracted via a least-squares fit and found to decrease in time as $\Gamma(t)\sim t^{-\delta_\mathrm{I} }$, with $\delta_\mathrm{I} =0.34(1)$  (insets of \Fig{defectseparation}a).
This confirms the scaling of the mean time to caustics within the error bounds.

To investigate the underlying spatial coarsening of the system, we recall the vortex structures shown in \Fig{realspacedefect} giving rise to a length scale in $\varphi_\mathrm{L}$. 
In \Fig{defectseparation}b, we depict the distribution spatial instanton separation in the system.
The resulting PDF exhibits an exponential fall-off
$P(r,t)\sim A(t) \exp[-r/\zeta(t)]$
with the mean separation increasing as 
$\expval{r} (t) = \int \dd{r} r\, P(r,t) \sim t^{\,\beta_\mathrm{I}}$
with exponent  
$\beta_\mathrm{I}=0.26(1)$,
cf.~the lower inset of \Fig{defectseparation}b, corroborating the results obtained in \cite{Schmied:2018osf.PhysRevA.99.033611}.
Hence, within the error bounds, the relation $\delta_\mathrm{c}=4\beta_V/3$ holds in the spatio-temporal scaling of the real-time instantons, which introduce a scale into the order parameter $F_\perp$.


\emph{Conclusions}.
Quenching a one-dimensional spin-1 Bose gas into the easy-plane phase 
leads to rich dynamics in the $F_x$-$F_y$ plane reflected in the fluctuations of the Larmor phase $\varphi_\mathrm{L}$.
Rare extreme rogue waves emerge from the disordered dynamics, which act as an effective random potential with a time varying correlation length $\ell_V(t)$ on the different components of the spinor gas. 
The time scale set by these events is found to scale as 
$t_\mathrm{c} \sim t^{\,4\beta_V/3} \sim t^{1/3}$,
which corroborates the coarsening dynamics of the spin correlations.
The focusing events give rise to real-time instantons, i.e., vortex structures in space and time in the Larmor phase, which in turn introduce the coarsening length scale found in the power spectrum of $F_\perp$.
These defects occur with algebraically decaying probability in time, reflecting, once more, a temporally coarsening time scale $\Gamma \sim t_\mathrm{c}^{-1}\sim t^{-1/3}$. 
Our results open a perspective on studying caustics \cite{Berry1977a,Berry1977b,Mumford2017a.JPB50.044005,Kirkby2019a.PhysRevResearch.1.033135,Kirkby2022a.PhysRevResearch.4.013105} leading to rogue waves \cite{Kaplan2002a.PhysRevLett.89.184103,Metzger2010a.PhysRevLett.105.020601,Metzger2013a.PhysRevLett.111.013901, Metzger2014a.PhysRevLett.112.203903, Degueldre2016a.NatPhys12.259} in multicomponent Bose condensates. 
Two exponents $\beta\equiv\beta_V\simeq\beta_\Lambda\simeq\beta_\mathrm{I}$ and $\delta\equiv\delta_\mathrm{c}\simeq\delta_\mathrm{I}$ emerge, reflecting a different algebraic growth in time, of the length and time scale, respectively, which are connected by $\delta=4\beta/3$.
The type of non-thermal fixed point observed in the multicomponent field model could bear interesting consequences for universal dynamics in the context of other systems, ranging from structure formation in the universe to non-linear hydrodynamics and microscopic physics.\\


\emph{Acknowledgements}.
The authors thank V.~Bagnato, J.~Berges, J.~Bloch, A.~Bulgac, K.~Geier, P.~Gro{\ss}e-Bley, P.~Heinen, M.~Karl, W.~Kirkby, A.~N.~Mikheev, R.~Miyar, J.~M.~Pawlowski, A.~Pi{\~n}eiro Ori\-oli, M.~Pr\"ufer, N.~Rasch, C.~M.~Schmied, T.~Simula, and S.~K.~Turitsyn  for discussions and collaboration on related topics. 
They acknowledge support 
by the ERC Advanced Grant EntangleGen (Project-ID 694561), 
by the German Research Foundation (Deutsche Forschungsgemeinschaft, DFG), through 
SFB 1225 ISOQUANT (Project-ID 273811115), 
grant GA677/10-1, 
and under Germany's Excellence Strategy -- EXC 2181/1 -- 390900948 (the Heidelberg STRUCTURES Excellence Cluster), 
and by the state of Baden-W{\"u}rttemberg through bwHPC and DFG through
grants INST 35/1134-1 FUGG, INST 35/1503-1 FUGG, INST 35/1597-1 FUGG, and 40/575-1 FUGG.

\begin{appendix}
\begin{center}
\textbf{APPENDIX}
\end{center}
\renewcommand{\theequation}{A\arabic{equation}}
\renewcommand{\thefigure}{A\arabic{figure}}
\setcounter{equation}{0}
\setcounter{figure}{0}
\setcounter{table}{0}
\makeatletter

\renewcommand{\theequation}{A\arabic{equation}}
\renewcommand{\thefigure}{A\arabic{figure}}
\setcounter{equation}{0}
\setcounter{figure}{0}
\setcounter{table}{0}
\makeatletter
\renewcommand{\theequation}{A\arabic{equation}}
\renewcommand{\thefigure}{A\arabic{figure}}
%

In the following we provide further details of the theory, numerical methodology, and of our results.

\section{Theory of the spin-1 Bose gas}
\label{app:SpinorBoseGas}
In this appendix, we briefly review the mean-field representation of the ground states of the spin-1 Bose-Einstein condensate on either side of the quantum phase transition between the polar and easy-plane (broken axi\-symmetric) phases, and discuss the semi-classical Truncated-Wigner methods used to simulate the non-equilibrium dynamics of the system.
%
\subsection{Polar to easy-plane phase transition}
\label{app:GroundState}
The contributions of the quadratic Zeeman effect and the terms proportional to $c_1$ in the Hamiltonian \eq{Spin1Hamiltonian} of the spin-1 Bose gas pose competing energy scales, which determine the nature of the ground state of the system, thus giving rise to various mean-field phases in the $q$-$c_1$ plane.
Our interest lies within the ferromagnetic regime where $c_1<0$, in the absence of the quadratic Zeeman shift, favours aligned spins, and where a second-order quantum phase transition controlled by $q$ divides the polar and easy-plane phases.
For 
$q>2n\abs{c_1}$
the system is in the polar phase, which shows no magnetization. 
Its mean-field ground state is given by 
\begin{align}
    \vb*{\Psi}_\mathrm{P} = e^{i\theta}\,\mqty(0 \\ 1 \\ 0)
    \,,
    \label{eq:PolarMFGS}
\end{align}
where $\theta$ is a global $U(1)$ phase of the condensate.
Conversely, for 
$0<q<2n\abs{c_1} \equiv 2\tilde{q}$,
the system is in the easy-plane phase, with the ground-state spinor field  
\begin{align}
    \vb*{\Psi}_{\mathrm{EP}} 
    = \frac{e^{i\theta}}{2}\,\mqty(e^{-i\varphi_\mathrm{L}/2}\sqrt{1-q/2\tilde{q}} \\ \sqrt{2+q/\tilde{q}} \\e^{i\varphi_\mathrm{L}/2}\sqrt{1-q/2\tilde{q}})
    \,.
\end{align}
In this phase, the system shows a magnetization transverse to the $F_{z}$-direction, with a complex order parameter 
$F_\perp=F_x+\mathrm{i}F_y$, %
exhibiting a total magnetization 
$\abs{F_\perp}=[1-q^2/(2\tilde{q})^2]^{1/2}$.
%
\begin{figure*}[t]
    \centering
    \includegraphics[width=0.9\textwidth]{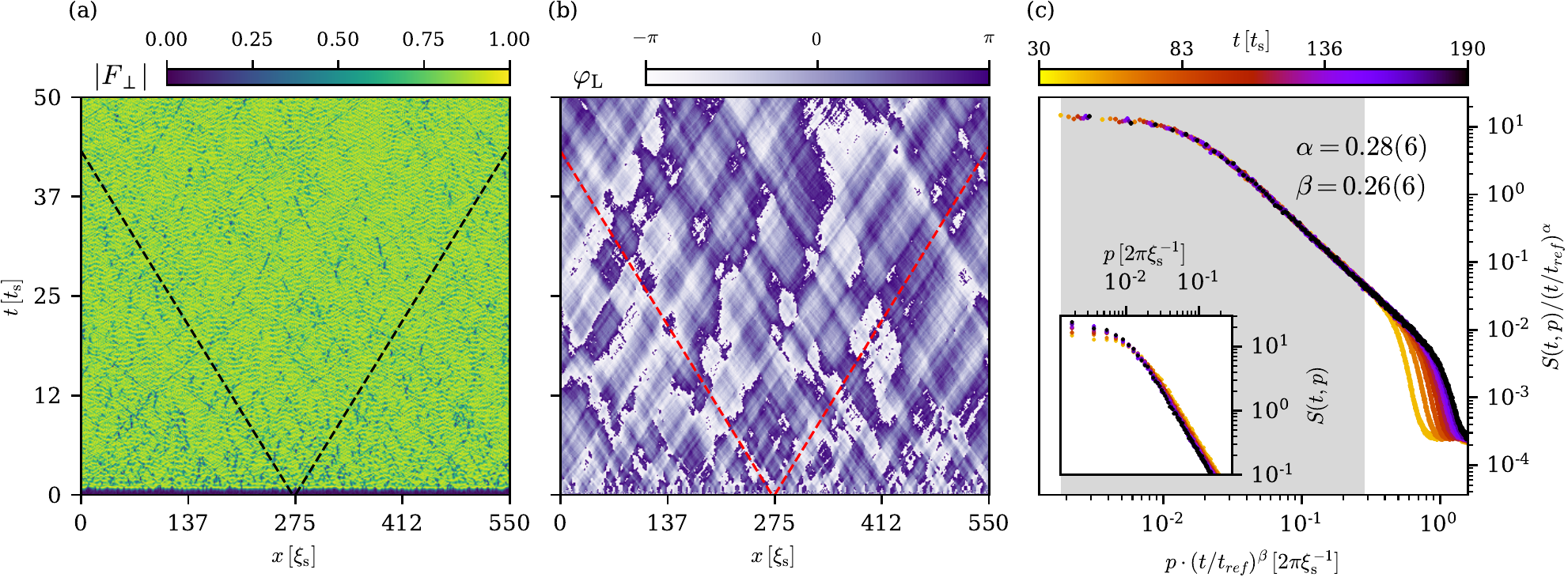}
    \caption{Time evolution (a) of the spin length $|F_{\perp}|$ of the transverse spin $F_{\perp}=F_{x}+\i F_{y}$  and (b) the Larmor phase $\varphi_\mathrm{L}=\mathrm{arg}(F_{\perp})$. The spin speed of sound $c_s=(n|c_1|/2M)^{1/2}$ is depicted by the black and red dashed lines.
    (c) The coarsening of spin-wave patterns seen mostly in the Larmor phase is reflected by the spatio-temporal scaling \eq{StructureFactorScaling} of the structure factor \eq{SFperp} with universal exponents $\beta=0.26(6)$ and $\alpha=0.28(6)\simeq d\beta$ in $d=1$ spatial dimension and universal scaling function $f_\mathrm{s}(p)$. The scaling exponents confirm, within the error bounds, the findings of \cite{Schmied:2018osf.PhysRevA.99.033611}.}
    \label{fig:FperpSpaceTime}
\end{figure*}
\subsection{Topology and order parameter}
\label{app:GroupTheory}
The order-parameter manifold of the system in the easy plane is given by 
$\mathcal{R}^\mathrm{EP}=$ SO$(2)_{F_z} \cross\, $U$(1)_\phi = $ S$^1_F \cross\,$U$(1)_\phi$,
where $F$ and $\phi$ denote the spin degree of freedom and global phase, respectively. 
The SO$(2)$ symmetry is to be understood with respect to the $F_z$-direction.
The bulk of the dynamics in the system takes place in spin space, hence we concentrate on the only non-trivial homotopy group for the spin part of $\mathcal{R}^\mathrm{EP}$ in one spatial dimension, which is given by 
$\pi_1($S$^1)=\mathbb{Z}$.
The Larmor phase $\varphi_\mathrm{L}$ corresponds to a Goldstone mode arising due to the broken SO$(2)_{F_z}$ symmetry and allows us to thus define the pure gauge field or topological current $j_\mu = \partial_\mu \varphi_\mathrm{L}$. 
Hence, we use this to define the topological charge $Q_{\mathrm{w}} = \int \dd{x} j_1^{\,\mathrm{w}}$.
We can therefore expect field configurations in the transverse spin degree of freedom to each correspond to a topologically distinct vacuum state with a well defined integer winding number $Q_{\mathrm{w}}$ of the Larmor phase, when going around the periodic system from $x=0$ to $x=\mathcal{L}$, the linear size of the system.

\subsection{Truncated Wigner simulations}
\label{app:TWA}
Within the far-from-equilibrium setting we focus on in this work, we simulate the dynamics of a cloud of $^{87}$Rb atoms in the $F=1$ hyperfine manifold using the Truncated-Wigner (TW) method \cite{Blakie2008a,Polkovnikov2010a.AnnPhys.8.1790}.
Within this spin manifold, the state of the system is quantified by the spinor  
$\vb*{\Psi} = (\Psi_1,\Psi_0,\Psi_{-1})^T$
formed by the complex scalar Bose fields describing the three magnetic components $m_\mathrm{F}=0,\pm 1$, and in terms of which the total particle density is given as
$n=\vb*{\Psi}^\dagger\!\cdot\! \vb*{\Psi}$.

We prepare the system to initially form a condensate in the zero-temperature mean-field ground state \eq{PolarMFGS} of the polar phase, which is characterized by a full macroscopic occupation of the $m_\mathrm{F}=0$ component 
$\psi_0(x)/\sqrt{n}=\expval{\Psi_0}/\sqrt{n}=1$,
while the side modes $m_\mathrm{F}=\pm1$ remain empty.
We add quantum noise to the Bogoliubov modes of the condensate following the prescription given in \cite{Schmied:2018osf.PhysRevA.99.033611}. 
The noise occupation of these Bogoliubov modes is crucial for the formation of the post-quench instabilities and the subsequent dynamics.

Starting from each thus prepared initial field configuration, the system is propagated by means of the classical field equations derived from the Hamiltonian \eq{Spin1Hamiltonian},
\begin{align}
    \mathrm{i} \partial_t \vb*{\Psi}(x,t) 
    = \biggl[&-\frac{\partial_x^2}{2M} + qf_z^2+ {c_0}n(x,t) 
    + {c_1} \vb*{F}(x,t)\cdot \vb*{f}\biggr]\vb*{\Psi}(x,t)
    \,.
    \label{eq:Spin1GPE}
\end{align}
Here, $M$ is the atomic mass, $q$ quantifies the quadratic Zeeman field, and U$(3)$-symmetric spin-independent interactions are described by the term $c_0n$.
Spin changing collisions are contained in the term ${c_1} \vb*{F}(x,t)\cdot \vb*{f}$, with
$\vb*{F}={\Psi}^\dagger_{m} \vb*{f}_{mn} {\Psi}_{n}$
and the $3\times3$ generator matrices $\vb*{f}=(f_x,f_y,f_z)$ of the $\mathfrak{so}(3)$ Lie algebra in the fundamental representation,
\begin{align}
  f_{x}=\mqty(0 & 1 & 0 \\ 1 & 0 & 1 \\ 0 & 1 & 0)\,,\quad
  f_{y}=\mqty(0 & -\i & 0 \\ \i & 0 & -\i \\ 0 & \i & 0)\,,\quad
  f_{z}=\mqty(1 & 0 & 0 \\ 0 & 0 & 0 \\ 0 & 0 & -1)\,.
  \label{eq:su3generators}
\end{align}

The physical parameters of the simulation such as mass and scattering lengths reflect realistic experimental values, yet we simulate a system with increased homogeneous density compared to the experiment \cite{Prufer:2018hto} and consider a purely homogeneous one-dimensional setting with no trapping potential.
We give spatial length in terms of the spin healing length 
$\xi_\mathrm{s}=(2Mn\abs{c_1})^{-1/2}$
and time in units of the characteristic spin-changing collision time 
$t_\mathrm{s} = 2\pi/(n\abs{c_1})$.
Furthermore, the field operators are normalized with respects to the total density 
$\Tilde{\Psi}_m=\Psi_m/\sqrt{n}$,
which results in a normalization of the spin vector as well 
$\tilde{\vb*{F}}=\vb*{F}/n$.
In the further discussion in the main text, the tilde is omitted and all values are to be understood as numerical values, unless explicitly stated otherwise.

Our one-dimensional numerical grid comprises $N=4096$ points with $3\cdot 10^6$ particles and is subject to periodic boundary conditions. 
The physical length of the grid corresponds to a length of $\mathcal{L}=220 \, \mu$m.
We choose the spin healing length to be given by $\xi_\mathrm{s}=8$ in lattice units, and our time stepping implies the spin collision time  to be $t_\mathrm{s}=696$ in numerical time units.
The propagation of \Eq{Spin1GPE} is done by means of a pseudo-spectral split-step Fourier method.
Finally, the relevant observables are evaluated by averaging the final field values over many samples of the initial conditions including noise (on the order of $\sim 10^3$ single TW runs).

Having implemented said simulation protocol on GPU clusters, the covered spectral range and evolution time are eventually limited by the size of the available memory and performance of the hardware, as the spectral method requires the data to be stored within the RAM of a single graphics card.

\section{Caustics and coarsening of correlations}
\setcounter{equation}{4}
\label{app:NumericalResults}
In this appendix, we discuss, in more detail, the coarsening evolution of the structure factor of the transverse spin $F_{\perp}$, as well as the caustics and their time-evolving statistics, which are appearing as a result of the spatio-temporally fluctuating spin excitations.
%
\begin{figure*}[t]
    \centering
    \includegraphics[width=0.9\textwidth]{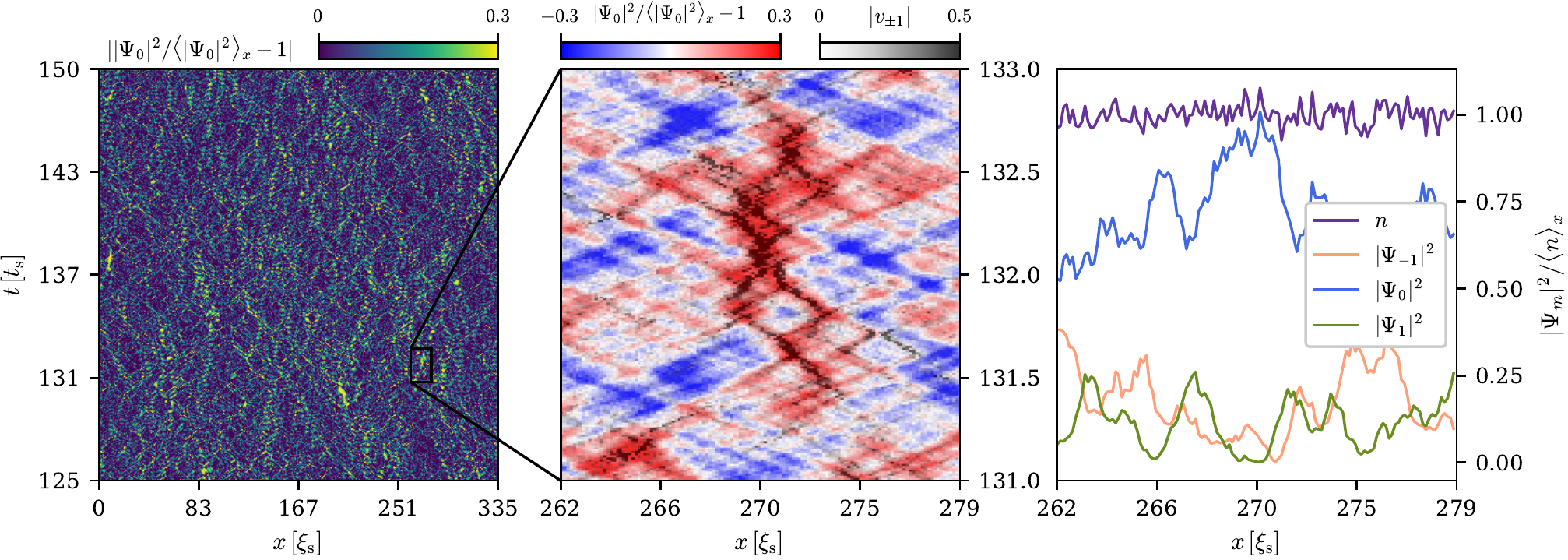}
    \caption{(a) Time evolution of the relative fluctuations $||\Psi_{0}|^2/\langle|\Psi_{0}|^2\rangle_x-1|$ of the $m_\mathrm{F}=0$ field amplitude around the spatial average $\langle|\Psi_{0}|^2\rangle_x$.
    Caustics show up as spikes in the relative density.
    (b) Magnification of a set of caustics demonstrating this spatio-temporal correlation. The deviation of the density $|\Psi_0|^2$ from its average is shown in blue to red colorscale.
    Overlayed on top are the velocity spikes $|\mathcm{v}_{\pm1}|$, demonstrating the correlation of a rise in density with the phase defects in the magnetic side modes.
    (c) The total density being approximately flat implies that the spikes seen in panel (a) correlate with dips in the magnetic side modes $m_\mathrm{F}=\pm1$. 
    }
    \label{fig:MagneticCaustics}
\end{figure*}

\subsection{Evolution of the transverse spin and Larmor phase}
\label{app:FtransverseEvolution}
Starting from the initially prepared condensate in the mean-field ground state of the polar phase, which has vanishing mean spin $\langle\mathbf{F}\rangle=0$, the short-time evolution is characterized by the appearance of instabilities in the transversal spin degree of freedom, causing a build-up of spatial structure  \cite{Schmied:2018osf.PhysRevA.99.033611}.
This transverse spin can be written, in a density-phase representation, as
$F_\perp \equiv F_{x}+\i F_{y} =\abs{F_\perp}\exp[\i\varphi_\mathrm{L}]$,
where $\varphi_{\mathrm{L}}=\varphi_{1}-\varphi_{-1}$ is known as the Larmor phase.
One observes the formation of patches of approximately equal order-parameter values, reflected mostly in the Larmor phase degree of freedom (see \Fig{FperpSpaceTime}b), as the spin length fluctuates weakly (cf.~green distribution in \Fig{FperpSpaceTime}a).

As reported in Ref.~\cite{Schmied:2018osf.PhysRevA.99.033611}, the patterns seen in the transverse spin, during the late-time evolution, cause the  structure factor 
\begin{align}
   S_{F_{\perp}}(t,p)=\langle F_{\perp}(t,p)^{\dagger}F_{\perp}(t,p)\rangle
   \label{eq:SFperp}
\end{align}
to scale in time and (momentum) space according to the universal form \eq{StructureFactorScaling}, with $\beta=0.26(6)$ and $\alpha=0.28(6)\simeq d\beta$, in $d=1$ spatial dimensions, cf.~\Fig{FperpSpaceTime}c.
This is understood to signal the approach of a non-thermal fixed point characterised by the quoted universal scaling exponents as well as the scaling function $f_\mathrm{s}(p)$.
Universality here means that, within a certain range of initial conditions and parameter values chosen, the time evolution leads to the same kind of scaling behavior in time and space, irrespective of the details of the initial condition and the details of the chosen parameter values.
We emphasise that the microscopic reason for the observed scaling exponents quoted above is unknown to date.

\subsection{Statistics of caustics in the spin-1 gas}
\label{app:Caustics}
As summarized in \Fig{caustic}, the patterns seen in the universal scaling evolution of the structure factor $S_{F_{\perp}}$, cf.~\App{FtransverseEvolution}, contain rogue-wave-like patterns in the phase gradients $\mathcm{v}_{m_\mathrm{F}} \sim \partial_x \varphi_{m_\mathrm{F}}$ of the magnetic components $m_\mathrm{F}=\pm1$.
These spikes in the superfluid velocity of the single components correlate spatially with peaks in the density $n_{0}$ of $m_\mathrm{F}=0$ component above a mean background density $\langle n_{0}\rangle$ attained in the long-time evolution, cf.~\Fig{MagneticCaustics}a, b.
These peaks are accompanied by corresponding dips in the densities of the magnetic side modes, $n_{\pm1}$, as the total density $n$ is, due to the much larger coupling $c_{0}\gg c_{1}$, to a good approximation constant in space and time, \Fig{MagneticCaustics}c.
In the following we analyse the spatial and temporal statistics of the occurrence of these caustics and relate it to the universal scaling seen in the transversal spin patterns, $F_{\perp}$.

Our results support the interpretation that a caustic in one of the magnetic sublevels is caused by the fluctuating potential, the respective other components effectively represent for its time evolution.
While the non-linear coupling present in the three-component system eventually gives rise to the dynamics seen, a separation into single modes evolving in a fluctuating background formed by the respective other ones, allows for a basic characterisation of the observed relation between the temporal and spatial scales.

An important question typically considered in the theory of caustics concerns the relation between the temporal and spatial scales characterizing the frequency of the occurrence of rogue waves and their mean separation in space, respectively.
To this end, one determines, starting from some random initial state, the mean time to first caustics on the basis of the spatio-temporal correlations of the random noise potential.

\begin{figure*}[t]
    \centering
    \includegraphics[width=0.9\textwidth]{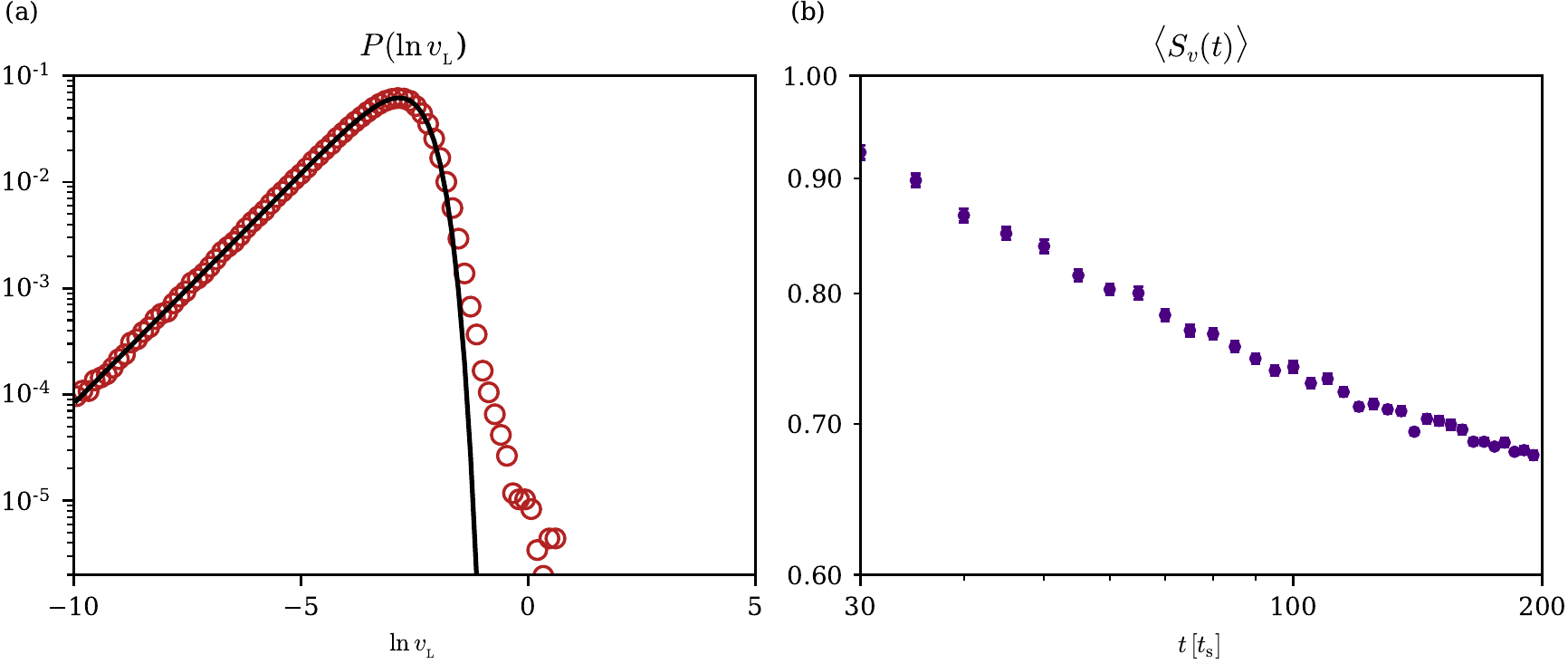}
    \caption{Statistics of caustics in the spin-1 gas. 
    (a) PDF of $\ln \mathcm{v}_{_\mathrm{L}}$ showing a Rayleigh exponential distribution with a heavy tail, in accordance with quantum dynamics of coherent wave packets \cite{Metzger2013a.PhysRevLett.111.013901}.
    (b) 
    Time evolution of the scintillation index averaged over all realizations on a double logarithmic scale. The scintillation index decays over time, confirming the decrease in the occurrence of rogue wave events.
    }
    \label{fig:scintindexandlogv}
\end{figure*}

Following the arguments of \cite{Kaplan2002a.PhysRevLett.89.184103,Metzger2010a.PhysRevLett.105.020601,Metzger2013a.PhysRevLett.111.013901,Metzger2014a.PhysRevLett.112.203903}, the time to first caustics is inferred from the solution of the classical equations of motion of point particles in the random potential $V$,
\begin{align}
    dx(t)/dt &= p_{x}(t)/m
    \,,& 
    dp_{x}(t)/dt
    &= -\partial V(x,t)/\partial x
    \label{eq:classicalppeom}
    \,,
\end{align}
i.e., integrating over time, from
\begin{align}
    x(t) &= x_{0}+\frac{p_{0}}{m}(t-t_{0}) -\int_{t_{0}}^{t}\mathrm{d}t'\int_{t_{0}}^{t'}\mathrm{d}t''\frac{\partial V(x,t'')}{\partial x}
    \,,
\end{align}
with $x_{0}=x(t_{0})$, $p_{0}=p_{x}(t_{0})$ defining the initial position and momentum.
This is equivalent to considering the motion of particles with large linear velocity $p_{y}/m$ in a weak two-dimensional random potential $V(x,y)$, in the paraxial approximation, $p_{y}\gg p_{x}$ \cite{Metzger2014a.PhysRevLett.112.203903}.
Choosing $x_{0}=p_{0}=0$, the mean squared variation of rays at time $t$ results as
\begin{align}
    \expval{x(t)^{2}} 
    &=
    \int_{t_{0}}^{t}\!\!\mathrm{d}t'\!\!\int_{t_{0}}^{t}\!\!\mathrm{d}\bar{t}'\!\!
    \int_{t_{0}}^{t'}\!\!\mathrm{d}t''\!\!\int_{t_{0}}^{\bar{t}'}\!\!\mathrm{d}\bar{t}''\!
    \left.\expval{\frac{\partial V(x,t'')}{\partial x}\frac{\partial V(x',\bar{t}'')}{\partial{x'}}}\right|_{x=x'}
    \!\!,
    \label{eq:xvarianceclassical}
\end{align}
where the average is taken over many realisations of the noise potential.

In the theory of caustic formation, the temporal correlations are often assumed to be Markovian, i.e., proportional to a delta-distribution in the relative time, 
\begin{align}
    C_{V}(\tau) 
    &= \expval{V(x,t)V(x,t+\tau)}
    = V_{0}^{2}(t)\tau_{0}\,\delta(\mathrm{\tau})
    \,,
    \label{eq:NoisePotCorrTime}
\end{align}
with some time constant $\tau_{0}$.
Furthermore, the spatial correlations of the potential typically show some Gaussian or exponentially fall-off, e.g.,
\begin{align}
    &C_{V}(r) 
    = \expval{V(x,t)V(x+r,t)}
    = V_{0}^{2}(t)\,\mathrm{e}^{-r/\ell_V(t)}
    \,.
    \label{eq:NoisePotCorrSpace}
\end{align}

Finally, the time to first caustics $t_\mathrm{c}$ is estimated by evaluating the noise correlator, integrate it over time, and demand that the variance at time $t=t_{0}+t_\mathrm{c}$ is on the order of the correlation length squared $\ell_V^{2}$ of the noise. 
This condition implies that caustics occur as the result of focusing of rays originating from a window around the focusing point the size of which corresponds to the scale on which the potential varies.
From \Eq{xvarianceclassical} one thus finds the scaling relation $t_\mathrm{c}\sim \ell_V^{4/3}$.

In the spin-1 Bose gas considered here, all particles belong to one of the three magnetic sublevels.
Specifically, the system follows the classical field equation \eq{GPE1}, with a potential term \eq{app:NoisePot} comprising the spin-spin coupling, which, together with the quadratic Zeeman shift $\sim q f_{z}^{2}$, breaks the U$(3)$ symmetry of the model by lifting the energy degeneracy and allowing for spin-changing collisions. 
Hence, in order to estimate the time to first caustics, the full three-component equation needs to be taken into account. 
The matrix potential $V$, due to $c_{1}\ll c_{0}$ and its non-linear dependence on the fluctuating fields, effectively takes the role of a weak noise which causes the spin-wave excitations to form caustics.

Analogously to the above introduced arguments of \cite{Kaplan2002a.PhysRevLett.89.184103,Metzger2010a.PhysRevLett.105.020601,Metzger2013a.PhysRevLett.111.013901, Metzger2014a.PhysRevLett.112.203903}, we estimate the time to first caustics in a semi-classical way.
The excitations leading to caustics are dominated by Goldstone-type excitations which redistribute particles within single magnetic sublevels while the total density $n$, subject to the density-density interactions $c_{0}n^{2}$ remains nearly constant.
The spikes observed in the velocities correlate with peaks in the $m_\mathrm{F}=0$ component, balanced by dips in $m_\mathrm{F}=\pm1$, recall \Fig{MagneticCaustics}.

That the formation of caustics in the system is characterized by the propagation of coherent waves in a random background is corroborated by the probability density function (PDF) of $\ln \mathcm{v}_{_\mathrm{L}}$ as shown in \Fig{scintindexandlogv}(a). 
It exhibits a Raleigh exponential shape with a heavy tail, in accordance with quantum dynamics of coherent wave packets \cite{Metzger2013a.PhysRevLett.111.013901}.
Note that, as before, we characterize the intensity of caustics by means of the gradient of the Larmor phase.
We have also computed the long-time evolution of the scintillation index \eq{scintillationindex}, averaged over many realisations, see \Fig{scintindexandlogv}(b). 
It decays in time, reflecting again the decrease in the density of rogue-wave events over time.


Hence, we need to estimate the time to first caustics from the time evolution of the fields capturing the three sublevels.
We consider a caustic at some time $t_{0}$ in sublevel $m$, which is described by a distribution of the deviation $\delta\Psi_{m}(x,t_{0})\equiv\Psi_{m}(x,t_{0})-\langle\Psi_{m}\rangle$ from a stationary mean value $\langle\Psi_{m}\rangle$, peaked around a position $x_{0}$, with width 
\begin{align}
	\langle [x-x_{0}]^{2}\rangle_{m,t_{0}}
	=\frac{\int \mathrm{d}x\,  (x-x_{0})^{2}|\delta\Psi_{m}(x,t_{0})|^{2}}{\int \mathrm{d}x\, |\delta\Psi_{m}(x,t_{0})|^{2}}
	\simeq \xi_\mathrm{s}^{2}
	\,
\end{align}
being on the order of the spin healing length, as can be inferred from \Fig{MagneticCaustics}.
The task is to estimate the temporal increase of the variance $\langle [x(t)-x_{0}]^{2}\rangle_{m}$ due to the evolution in the noisy potential formed by the other magnetic components.
For this, we need to estimate the time evolution of the field starting from the caustic peak $\delta\Psi_{m}(x,t_{0})$.

This evolution is governed by the Hamiltonian \eq{Spin1Hamiltonian}, which can be split into a Bogoliubov mean-field (MF) part and the rest, $H=H_\mathrm{MF}+\delta H_{V}$, where the Bogoliubov MF term, which is at most quadratic in the fields $\Psi_{m}$, gives rise to a coherent background evolution.
As we can neglect, to a good approximation, Bogoliubov fluctuations of the total density $n$, the main contribution will arise from the gapless spin-wave excitations present in the easy-plane phase.
Furthermore, as the width of the caustic peak is on the order of the spin healing length, wave numbers contributing to the packet are $k\lesssim k_{\xi_\mathrm{s}}\sim 1/\xi_\mathrm{s}$.
Hence, the evolution with $H_\mathrm{MF}$ causes the wave packet to (split and) move, without dispersing, at the speed $c_\mathrm{s}=(n|c_{1}|/2M)^{1/2}$,
\begin{align}
    U_\mathrm{MF}^{\dagger}(t,t_{0})\delta\Psi_{m}(x,t_{0}) U_\mathrm{MF}(t,t_{0})
    \simeq \delta\Psi_m(x-c_\mathrm{s}[t-t_{0}],t_{0})\phi_{m}(x,t)
    \,.
    \label{eq:MFEvol}
\end{align}
Here $U_\mathrm{MF}(t,t_{0})= \exp(-\i \int_{t_{0}}^t \mathrm{d}t' H_\mathrm{MF})$, and we neglect possible weak effects from dispersion in higher wave numbers.
While the time evolution shifts the position of the wave packet, it in general also involves fast phase oscillations with a frequency on the order of $\sim\omega(k_{\xi_\mathrm{s}})$, which are taken into account by the, not further specified, multiplicative factor $\phi_{m}(x,t)$ which takes the form of a complex oscillating function of norm $|\phi_{m}(x,t)|\lesssim1$, cf.~\Fig{MagneticCaustics}.

Besides this coherent propagation of the packets with the speed of sound, the wave packet spreads out due to the motion in the noisy background potential
\begin{align}
    V(x,t) = c_1 \mqty(F_z & F_\perp^{*} & 0 \\
                    F_\perp & 0 & F_\perp^{*} \\
                    0 & F_\perp & -F_z)(x,t)
    \label{eq:app:NoisePot}
\end{align}
which enters the interaction Hamiltonian $H_{V}(t)=\int\mathrm{d}x\,\Psi_{m}^{\dagger}(x)V_{mn}(x,t)\Psi_{n}(x)$ and thus the beyond-MF part $\delta H_{V}=H_{V}-H_{V,\mathrm{MF}}$.
Note that we neglect beyond-MF contributions from the density-density interactions $\sim c_{0}n^{2}$ and that the potential $V$ is taken to represent a time-varying background potential despite the fact that the field operators $\Psi_{m}$ are evaluated at the fixed initial time $t_{0}$.
The resulting beyond-MF Hamiltonian encodes the noisy background, which fluctuates on lengths scales set by the fluctuations of $F_{z}$ and $F_{\perp}$, cf.~\Eq{app:NoisePot}.  
As is seen in our simulations, cf.~\Fig{FperpSpaceTime}, this length scale is the time-dependent infrared scale $\ell_{\Lambda}(t)\sim t^{\,\beta}$, which characterizes the size of the coarsening patterns in the Larmor phase and, equivalently, the infrared wave number $k_{\Lambda}\sim\ell_{\Lambda}^{-1}$ marking the onset of the plateau in the structure factor $S_{F_{\perp}}(k,t)$ in \Fig{FperpSpaceTime}c.
\begin{figure*}[t]
    \centering
    \includegraphics[width=0.85\textwidth]{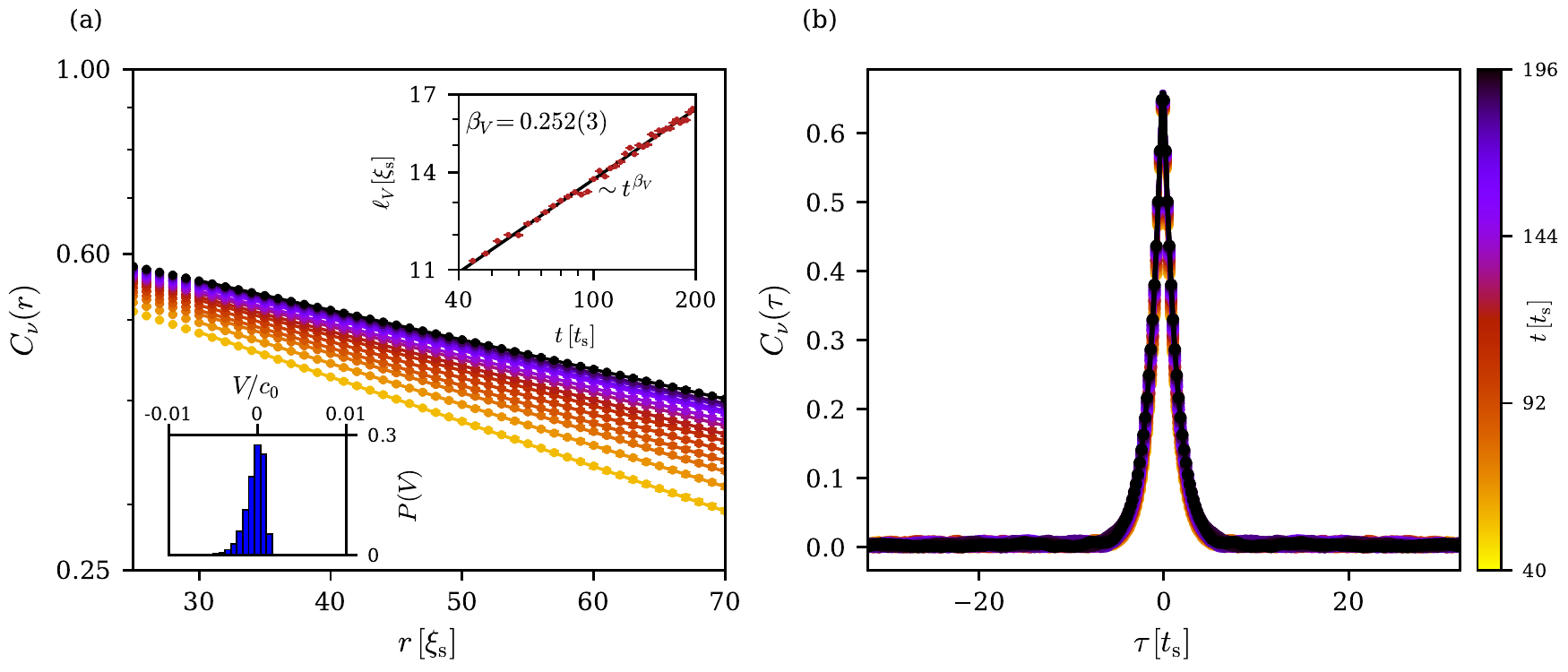}
    \caption{(a) Correlation function $C_{V}(r)=\expval{\int\dd{x} V(x,t)V(x+r,t)}$ of the random potential term in \Eq{app:NoisePot}. 
    $C_{V}(r)$ takes the form of an exponential function $\exp(-r/\ell_V)$ with a time varying characteristic correlation length $\ell_V \sim t^{\,\beta_{V}}$, with $\beta_{V} = 0.252(3)\simeq\beta$ (upper inset).
    The lower inset shows the potential is weak in comparison to density fluctuations. 
    (b) Temporal correlation function $C_V(\tau) = \expval{\int \dd{x} V(x,t)V(x,t+\tau)}$. The correlation shows a fast decaying Lorentzian with constant correlation time $\ell_\tau$.}
    \label{fig:correlationlength}
\end{figure*}

As $\ell_{\Lambda}\gg\xi_\mathrm{s}$, we can Taylor expand the noise potential at the position of the caustic peak to first order around a constant background,
\begin{align}
    V(x,t) = V(x_{0},t) + \left.\frac{\partial V}{\partial x}\right|_{x=x_{0}}(x-x_{0})+\dots
                    \,,
    \label{eq:PotTaylor}
\end{align}
and neglect the constant term, which causes the MF evolution to be corrected essentially to include a non-zero width of the dispersion.

The time-evolution operator $U(t,t_{0})= \exp(-\i \int_{t_{0}}^t \mathrm{d}t' H)$ can be split into a fast mean-field part and a slow evolution caused by $V$,
\begin{align}
  &U(t,t_{0})
  =U_\mathrm{MF}(t,t_{0})
  \nonumber\\
  &\ \times  \exp[-\i \int_{t_{0}}^t \mathrm{d}t' U_\mathrm{MF}^{\dagger}(t'-t_{0})\delta H_{V}(t')U_\mathrm{MF}(t'-t_{0})]
  \,.
  \label{eq:USplitting}
\end{align}
Combining the above expressions, we can now calculate the approximate broadening of the travelling wave packet due to the external noise $V$,
\begin{widetext}
\begin{align}
   \left< \left[x-c_\mathrm{s}(t-t_{0})-x_{0}\right]^{2}\right>_{m,t}
   &\simeq\mathcal{N}_{m}^{-1}\int \mathrm{d}x\,  [x-c_\mathrm{s}(t-t_{0})-x_{0}]^{2}
   \Bigg\{\
   |\delta\Psi_{m}(x-c_\mathrm{s}[t-t_{0}],t_{0})|^{2}
   \nonumber\\
   &\quad+\delta\Psi^{*}_{m}(x-c_\mathrm{s}[t-t_{0}],t_{0})
   \int\mathrm{d}t'\,\mathrm{d}t''\,
   \left.\left<\frac{\partial V_{ml}(x',t')}{\partial x'}\frac{\partial V_{ln}(x'',t'')}{\partial x''}\right>_{V}\right|_{x'=x''=x_{0}}
   \delta\Psi_{n}(x-c_\mathrm{s}[t-t_{0}],t_{0})
   \nonumber\\
   &\qquad\ \  \times
   \left[x-c_\mathrm{s}(t-t')-x_{0}\right]\left[x-c_\mathrm{s}(t-t'')-x_{0}\right]
   \Bigg\}
   \,,
   \label{eq:BroadeningCausticPeak}	
\end{align}
where $\mathcal{N}_{m}=\int \mathrm{d}x\, |\delta\Psi_{m}(x,t_{0})|^{2}$ is a normalisation, the mean value $\langle\cdots\rangle_{V}$ denotes averaging over the noise potential and we have dropped terms linear in $V$, as they vanish when taking this average.
We have also neglected any fast rotating phases, which will play a role in any single realisation of the potential but are expected to average out in the mean.
Note that the covariance of the noise involves a matrix product of the potential, which takes the form
\begin{align}
    C_{V}(r) 
    &= \expval{V(x,t)V(x+r,t)}_{V}
    \nonumber \\[0.2cm]
    & = \Bigg<\ \mqty(
    F_{z}(x)^{*}F_{z}(x+r)+F_{\perp}(x)^{*}F_{\perp}(x+r)&0 &F_{\perp}(x)^{*}F_{\perp}(x+r)^{*}\\
    0 &  F_{\perp}(x)F_{\perp}(x+r)^{*}+\text{c.c.} & 0 \\[0.2cm]
    F_{\perp}(x)F_{\perp}(x+r) & 0 & F_{z}(x)^{*}F_{z}(x+r)+F_{\perp}(x)F_{\perp}(x+r)^{*}
    )\ \Bigg>_{V}
    \,.
    \label{eq:NoisePotCorr}
\end{align}
\end{widetext}
On average, the off-diagonal elements of this matrix vanish due to the O$(2)$ symmetry of the spin configuration in the easy plane, recall the circle-shape histogram shown in the upper panel of \Fig{realspacedefect}d.
At the same time, our numerical simulations show that the diagonal elements, to a good approximation, exhibit correlations of the form given in \eq{NoisePotCorrSpace}, with $V_{0}^{2}$ replaced by the respective matrix elements $(V^{2}_{0})_{mn}\equiv V_{0,m}^{2}\delta_{mn}$, cf.~\Fig{correlationlength}a. 
We find that the fluctuations of the Larmor phase and thus of $F_{\perp}$ dominate the correlations while those of $F_{z}$ can be neglected.
The temporal correlations show a nearly Markovian character \eq{NoisePotCorrTime}, as is seen in \Fig{correlationlength}b, where we fit a Lorentzian to the $\tau$-dependence,
\begin{align}
    C_{V}(\tau)
    = \frac{V_{0}^{2}(t)}{1+[c_\mathrm{s}\tau/\ell_\mathrm{\tau}(t)]^{2}}
    \,,
    \label{eq:NoisePotCorrLorentzTime}
\end{align}
with $\ell_\mathrm{\tau}(t)$ constant in time.

The near-Markovian character of the temporal correlations \eq{NoisePotCorrTime} allows us to integrate over $t''$, which thereby is set equal to $t'$.
Finally, the spatial averaging done by integrating over $x$ evaluates \eq{BroadeningCausticPeak} to become
\begin{align}
   \left< \left[x-c_\mathrm{s}(t-t_{0})-x_{0}\right]^{2}\right>_{m,t}
   &\simeq
  \xi_\mathrm{s}^{2}\left\{1+\frac{\pi}{3}V_{0,m}^{2}
  \frac{\ell_{\tau}c_\mathrm{s}t^{3}}{\ell_V^{2}} \right\}
  \label{eq:BroadeningCausticPeakResult}
  \,.
\end{align}
For times $c_\mathrm{s}t\gg\xi_\mathrm{s}$, the term $\sim t^{3}$ dominates the width. 
Hence, as discussed above for the case of classical particles, demanding that the variance \eq{BroadeningCausticPeakResult}, at the time $t=t_\mathrm{c}$ to first caustics, is on the order of $\ell_V^{2}$, one obtains again the scaling relation for the mean time to first caustics,
\begin{align}
    t_\mathrm{c} 
    &\sim \left(V_{0,m}^{2}\ell_{\tau}c_\mathrm{s}\right)^{-1/3}\ell_V^{4/3}
    \,.
    \label{eq:tcscalinginellc}
\end{align}
Beyond caustics in classical particle trajectories \cite{Kaplan2002a.PhysRevLett.89.184103,Metzger2010a.PhysRevLett.105.020601,Metzger2013a.PhysRevLett.111.013901, Metzger2014a.PhysRevLett.112.203903}, the scaling relation \eq{tcscalinginellc} is known from the theory of caustics in quenched spin chains \cite{Kirkby2019a.PhysRevResearch.1.033135} and quantum many-body systems \cite{Kirkby2022a.PhysRevResearch.4.013105}.
\begin{figure*}[t]
    \centering
    \includegraphics[width=0.9\textwidth]{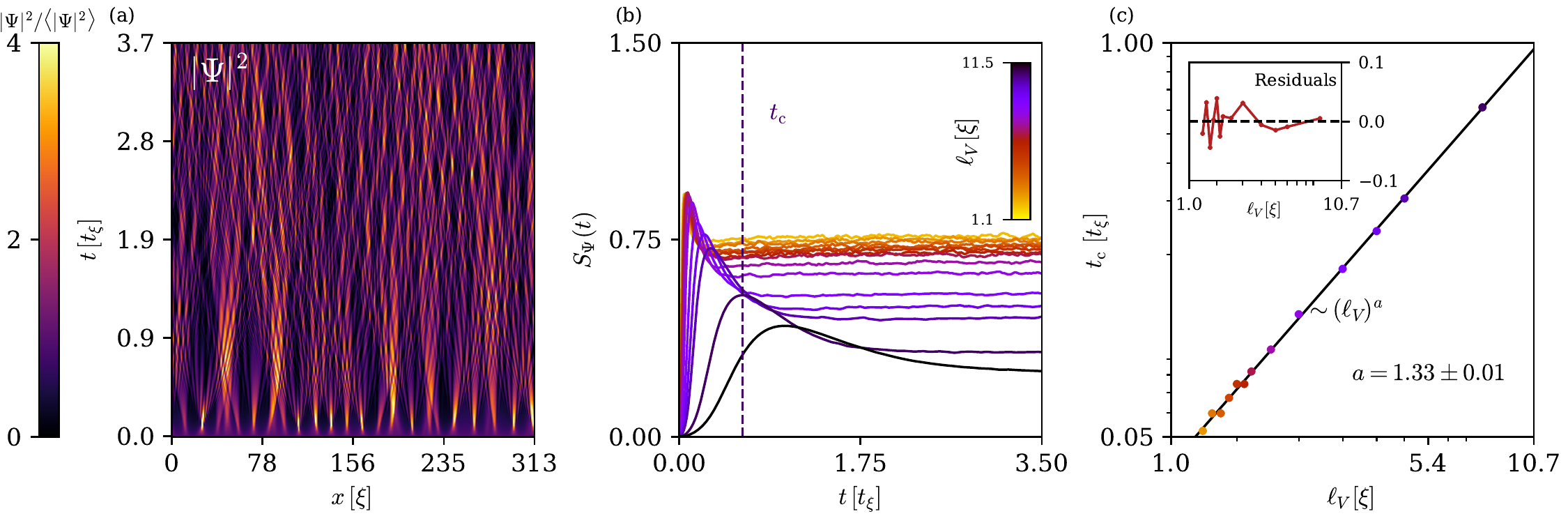}
    \caption{Caustics in one-dimensional GPE. 
    (a) Time evolution of the density $|\psi|^2$ of a single-component BEC in a random potential with $\ell_V=3 \, \xi$. Caustics form after a time $t_\mathrm{c}(\ell_V)$.
    (b) The scintillation index $S_\psi (t) = \langle{|\psi|^4}\rangle_x/\langle{|\psi|^2}\rangle_x^2-1$ shows the formation of caustics and the subsequent saturation for various correlation lengths.
    Each curve is a result of an averaging over $\sim 10^2$ runs. 
    (c) Dependency of the mean time to first caustics on the correlation length of the random potential. Using a least squares fit we extract the scaling of the mean time to caustics $t_\mathrm{c}\sim (\ell_V)^a$, with $a=1.33(1)$.
    The inset shows the unweighted deviation of the fit from the data.}
    \label{fig:1cGPECaustics}
\end{figure*}

\subsection{Caustics in a single-component Gross-Pitaevskii system}
\label{app:GPEcaustics}
We have reproduced the scaling \eq{tcscalinginellc} of $t_\mathrm{c}$ relative to $\ell_V$  by evaluating the non-linear dynamics according to a single-component GPE in a random-noise potential, with correlation length $\ell_V$ chosen constant in time.
For this, we have simulated a single-component GPE with non-linearity $g$ and a random potential corresponding to the parameters of \Eq{Spin1GPE},
\begin{align}
    \mathrm{i} \partial_t \psi = \left[-\frac{1}{2M}\partial_x^2 + g\abs{\psi}^2 + V(x)\right]\psi
    \,.
\end{align}
The initial condition was chosen to be a constant field with homogeneous density $\rho = 3190\xi^{-1}$, and $V(x)$ a randomly fluctuating and correlated weak potential with a correlation length $\ell_V$, which is chosen to be constant in time.
The simulation corresponds to a physical system length of $\mathcal{L}=220\, \mu$m $= 313\xi$, with $\xi=(2M \rho g)^{-1/2}$ and an integration time of $t_{\mathrm{max}}=14 \, $ms $ = 3.7 t_\xi $, with $t_\xi = 2\pi(2M\rho g)^{-1}$.
The ensuing dynamics of the system gives rise to focusing events in the gas as seen in \Fig{1cGPECaustics}a.
For each chosen correlation length, we have performed $\sim 10^2$ runs and calculated the density scintillation index (see \Fig{1cGPECaustics}b)
\begin{align}
    S_\psi = {\expval{|\psi|^4}_x}/{\expval{|\psi|^2}_x^2}-1
    \,.
\end{align}
The mean time to first caustics $t_\mathrm{c}$ is then found as the first maximum of the scintillation index and poses a noise dependent characteristic time scale upon which caustics appear. 
The scaling behavior of $t_\mathrm{c}$ was then obtained using a least-squares fit of a power law $\ell_V^{\,a}$ with $a=1.33(1)$, in accordance with the relation \eq{BroadeningCausticPeakResult}.
%

\subsection{Temporal scaling of $\ell_V(t)$ and $t_\mathrm{c}(t)$}
\label{app:ScalingCausticCorrelations}
The self-similar scaling of the $F_\perp$ structure factor \eq{StructureFactorScaling} is reflected in the scaling behavior of the correlation length of spin fluctuations in the system, see \Fig{FperpSpaceTime} and Fig.~2 in \cite{Schmied:2018osf.PhysRevA.99.033611}.
When extracting the spatial correlation function $C_{V}(r,\tau)$, see \App{Caustics},
we determined, at each point of time, the mean length scale $\ell_V(t)$ upon which the correlation function decays using an exponential dependence \eq{NoisePotCorrSpace} on $r$, see \Fig{correlationlength}a.
We obtain a power-law scaling of the correlation length $\ell_V(t) \sim t^{\,\beta_{V}}$, with $\beta_{V}=0.252(3)$, consistent within errors with $\beta$ obtained from the structure factor, \Fig{FperpSpaceTime}.

Inserting this scaling of $\ell_V$ into the relation \eq{BroadeningCausticPeakResult} one finds that the mean time to first caustics scales in time as 
\begin{align}
    t_\mathrm{c}(t)\sim t^{4\beta_V/3}\sim t^{1/3}
    \,.
\end{align}
This constitutes a separate scaling law for the time scale $t_\mathrm{c}$ characterizing rogue waves due to caustics, exhibiting a different exponent as compared to $\ell_V(t)\sim t^{1/4}$.

\subsection{Scaling of Larmor phase fluctuations and winding number}
\label{app:WindingNumber}
With the help of the topological current $j_0 = \partial_t \varphi_\mathrm{L}$, we identify textures in the Larmor phase and their propagation through the condensate. 
As we demonstrate in the main text, caustics in these textures give rise to instantons, vortex-type defects in space and time.
To quantify the spatial and temporal correlations of these defects, we identify their positions by weighting the current with dips in the spin length, which results in the topological current
\begin{align}
    \mathcal{J}(x,t) 
    = |\partial_x\varphi_\mathrm{L}(x,t)| \cdot \left[\expval{|F_\perp|}_x - |F_\perp(x,t)|\right]
    \,.
\end{align}
where $\expval{\cdots}_x$ denotes the spatial average.

In order to calculate a characteristic length scale at each time, we utilize peak detection algorithms on the topological current $\mathcal{J}$, setting the peak values to be in the upper 10th percentile of its distribution function, ensuring that we detect extreme events, cf.~the upper right inset in \Fig{defectseparation}a.
We compute the distance between a defect and its next neighbor, and average over many realizations of the system, thereby finding the probability distribution $P(r,t)$ at each time of finding the distance $r$ from a defect to its next neighbor (\Fig{defectseparation}a).
At each time, the probability distribution of defect separation is found to obey, to a good approximation, an exponential form 
$P(r,t)\sim A(t) \exp(-r/\zeta(t))$
with a characteristic mean separation scale $\expval{r}(t) = \int \dd{r} r \cdot P(r,t)$ that varies in time.
The lower left inset of \Fig{defectseparation}a shows that $\expval{r}$ increases in time according to a power law 
$\expval{r}\sim t^{\beta_\mathrm{I}}$
with an exponent  
$\beta_\mathrm{I}=0.26(1)$ 
\footnote{
Here we introduce a notation which distinguishes the exponent $\beta_\mathrm{I}$ characterizing the length scale associated with instantons as compared to the general $\beta$ characterizing the coarsening \eq{StructureFactorScaling} of the structure factor, in order to not mingle our results upfront.
As it turns out, $\beta_\mathrm{I}$ and $\beta$ can not be distinguished within our numerical accuracy and are expected to b equivalent to each other.}.

To analyse the temporal frequency of the instanton events, we calculate the windowed Fourier transform (also short-time Fourier transform, STFT) of the winding number and average it over all runs.
The winding-number evolution for a single run is shown in the lower panel of \Fig{defectseparation}b.
At each point in time, we obtain a spectrum, which falls off steeply in frequency and, at low $\omega$, can be approximated by a Gaussian distribution, as seen in the lower inset of \Fig{defectseparation}b.
The typical frequency width $\Gamma$ of the Gaussian is extracted via a least-squares fit, and is found to vary in time and follow a power law as well. 
\end{appendix}


\bibliographystyle{apsrev4-1}

\end{document}